# Quantum-elevated Chiral Discrimination for Bio-molecules

Yiquan Yang[1,2,3], Xiaolong Hu[1,2,3], Wei Du[1,2,3], Shuhe Wu[1,2,3], Peiyu Yang[1,2,3], Guzhi Bao[1,2,3]*, Weiping Zhang[1,2,3,4,5]†

**Affiliations:**

[1]School of Physics and Astronomy, Shanghai Jiao Tong University; Shanghai, 200240, China.

[2]Tsung-Dao Lee Institute, Shanghai Jiao Tong University; Shanghai, 200240, China.

[3]Shanghai Branch, Hefei National Laboratory; Shanghai, 201315, China.

[4]Shanghai Research Center for Quantum Sciences; Shanghai, 201315, China.

[5]Collaborative Innovation Center of Extreme Optics, Shanxi University; Shanxi, 030006, China.

*Corresponding author. Email: guzhibao@sjtu.edu.cn

†Corresponding author. Email: wpz@sjtu.edu.cn

**Abstract:** Chiral discrimination of enantiomeric biomolecules is vital in chemistry, biology, and medicine. Conventional methods, relying on circularly polarized light, face weak chiroptical signals and potential photodamage. Despite extensive efforts to improve sensitivity under low-photon exposure, classical chiral probes remain fundamentally bound by the shot-noise limit due to quantum fluctuations. To beat these limitations, we demonstrate quantum-elevated chiral discrimination using continuous-variable polarization-entangled states as moderate-photon-flux, high-sensitivity, quantum-noise-squeezed chiral probes. We achieve a 5 dB improvement beyond the SNL in distinguishing L- and D-amino acids in liquid phase. This non-destructive, biocompatible protocol enables high-sensitivity chiral analysis, with broad implications for drug development, biochemical research, environmental monitoring, and asymmetric synthesis.

**Teaser:** Quantum-enhanced discrimination of molecular chirality is achieved using continuous-variable polarization-entangled light.

## INTRODUCTION

Chirality, a geometric property associated with the breaking of mirror symmetry, is particularly significant in biological molecules, which appear as non-superimposable mirror images known as enantiomers (*1*). Based on their handedness, chiral molecules can be classified as either left-handed (L) or right-handed (D) enantiomers. Enantiomers often exhibit markedly different biological activities. In particular, biological systems selectively utilize specific enantiomers, L-amino acids and D-sugars, as essential building blocks. This stereospecificity extends to pharmacology, where single-enantiomer drugs often exhibit superior efficacy over their racemic mixture (*2*, *3*). More critically, while one enantiomer may serve as a potent therapeutic agent, its mirror image can provoke severe adverse effects. Sensitive enantiomeric discrimination is therefore essential for biological research, drug development, and disease diagnosis (*4*, *5*).

Chiroptical methods are the dominant approach for enantiomeric discrimination, such as optical rotatory dispersion (ORD) (*6*, *7*), electronic or vibronic circular dichroism (*8–11*), and Raman optical activity (*12*, *13*). However, their reliance on magnetic dipole (or electric quadrupole) interactions results in weak chiral responses, necessitating long measurement times and large sample volumes. Moreover, as the transition energy bands of chiral molecules predominantly lie in the ultraviolet region, prolonged exposure may affect the biological and chemical activity of biomolecules (*14–16*). To overcome above limitations, considerable efforts over the past decade have focused on developing more sensitive chiral discrimination techniques under low optical damage. Advances in nano-photonics and high-finesse cavity fabrication have enabled significant enhancement of chiral light-matter interactions in the visible spectrum (*17–27*), with reduced optical damage. However, certain engineered nanomaterials may introduce excess background noise (*28*, *29*), potentially contaminating chiral signals and compromising analysis. Beyond engineering interaction platforms, sculpting the spatial and temporal structure of optical fields provides an alternative technological route to enhancing chiral matter-light interactions (*30–39*). Despite these methodological advancements, sensitivity of chiroptical methods remains fundamentally constrained by the shot noise limit (SNL), $1/\sqrt{N}$ with sensing photon number $N$, due to quantum fluctuations in the optical field. While increasing optical intensity can improve sensitivity, it also raises the risk of inducing optical damage, creating a trade-off between improved sensitivity and sample safety.

Recent advances in quantum optical metrology provide promising techniques to address this challenge. As a proof-of-principle, the photonic N00N state has been used as a quantum probe to demonstrate quantum enhancement in applications such as phase sensing (*40–42*), holography (*43*), Earth rotation measurements (*44*), and ORD experiments (*45*). However, so far, experimentally available N00N states are constrained to only a few photons, making their overall sensitivity far from that of conventional coherent probes. With continued advancements, high-brightness and high-sensitivity continuous-variable (CV) quantum squeezing is ready to become a versatile quantum technology, enabling breakthroughs across multiple disciplines. Its applications now span gravitational wave detection (*46*, *47*), magnetometry (*48–50*), nonlinear microscopy (*51*), SU(1,1) interferometer (*52–57*), atomic force microscopy (*58*). Further applications of quantum squeezing to sub-shot-noise imaging (*59–61*) and quantum illumination (*62–65*) have been demonstrated as well. Meanwhile, for biological applications relying on polarization measurements (*45*, *66–69*), CV polarization-entangled beams (*70*, *71*) are anticipated to offer superior sensitivity compared to conventional methods.

In this paper, we demonstrate the quantum-elevated chiral discrimination with CV polarization entanglement, enabling moderate photon flux and high sensitivity through quantum noise suppression. As illustrated in Fig. 1, we first generate two orthogonally polarized two-mode squeezed states (TMSS) using a pair of parametric amplifiers (PAs) via four-wave mixing processes of $^{85}$Rb atomic ensembles. These TMSS are coherently mixed at a polarization beam splitter (PBS) to construct the CV polarization-entangled state, which serves as a quantum probe for enantiomer discrimination via molecular optical activity. The chiral matter-photon interaction induces differential phase shifts between left- and right-circularly polarized modes, leading to a rotation of the linear polarization upon transmission through the chiral medium. Since the rotation angle is tiny for the extremely weak interaction, the resulting signal is buried in photon shot noise with a coherent light probe. By employing a CV polarization-entangled quantum probe, we demonstrate SNL-breaking quantum-elevated detection where the chiral signal emerges clearly from the squeezed noise floor. This sophisticated protocol allows for label-free and sample-safe chiral discrimination in a bio-friendly manner, with potential applications in bio-molecular analysis, drug activity, toxicity assessment, and biological process monitoring.

## RESULTS

### Entanglement-enabled Chiral Discrimination

Chiral enantiomers exhibit optical activity due to the interference between electric and magnetic dipole transitions, leading to distinct interaction with left- and right-handed circularly polarized light. In the semi-classical framework, the natural optical polarization rotation $\theta$ is determined by the dynamic molecular property $G_{\alpha\alpha}$ (*72*), which can be expressed as:

$$G_{\alpha\alpha} \propto \Sigma_{i\neq g} \operatorname{Re}\left(\langle g \,|\, \boldsymbol{\mu}_{\alpha} \,|\, i\rangle\langle i \,|\, \boldsymbol{m}_{\alpha} \,|\, g\rangle\right), \tag{1}$$

where the sum runs over all eigenstates for chiral molecules except for the ground state $|\,g\rangle$, $\boldsymbol{\mu}$ and $\boldsymbol{m}$ represent the induced electric and magnetic dipole moments, respectively. The index $\alpha = \{x, y, z\}$ denotes the spatial coordinate axis. The parallel and antiparallel orientations of electric and magnetic dipole moments induce clockwise and counterclockwise optical polarization rotations, respectively, enabling the discrimination between L- and D-enantiomers.

Conventionally, polarization rotation measurements are performed using coherent laser light, whose photon number follows a Poisson distribution. The associated quantum fluctuations introduce uncertainty in the polarization direction, limiting the sensitivity to the SNL (*73*, *74*). Since SNL is inversely proportional to the square root of the average photon number, increasing laser intensity can improve the sensitivity directly. However, high-intensity laser input can lead to unavoidable photon-induced damage, especially for *in vivo* testing, thus posing a trade-off between sensitivity improvement and sample safety that presents a critical challenge.

To overcome this, we propose a quantum protocol for chiral discrimination that mitigates this trade-off by enhancing sensitivity without increasing optical power. The key lies in suppressing the quantum fluctuations of the chiral probe. As shown in Fig. 1, quantum chiral discrimination consists of four main stages: CV entangled source, chiral matter-photon interaction, polarization detection and data processing. Two-mode squeezed states are generated via PAs, characterized by SU(1,1) matrix, with the input-output operator transformation following:

$$\begin{bmatrix} \hat{a}_{1,k} \\ \hat{b}_{1,k}^{\dagger} \end{bmatrix} = \begin{bmatrix} G & g \\ g & G \end{bmatrix} \begin{bmatrix} \hat{a}_{0,k} \\ \hat{b}_{0,k}^{\dagger} \end{bmatrix}, \tag{2}$$

The operators $\hat{a}_{i,k}$ and $\hat{b}^{\dagger}_{i,k}$ are the annihilation and creation ones of the input $(i=0)$ and output $(i=1)$ modes of the PAs, with $k=H,V$ denoting the horizontal and vertical polarization, respectively. The gain of PAs is characterized by $G=\cosh(r)$ and $g=\sinh(r)$, where $r$ is the squeezing parameter. The matrix elements satisfy the identity $G^2-g^2=1$, ensuring consistency with bosonic commutation relations. By superimposing a pair of two-mode squeezed states, one H-polarized and one V-polarized, on a PBS, we generate CV polarization entanglement with H-H and V-V correlation between modes $a$ and $b$. The intensity difference operators between the horizontal and vertical polarization components for the modes $a$ and $b$ are defined as $\hat{N}_{aj,-}=\hat{a}^{\dagger}_{j,H}\hat{a}_{j,H}-\hat{a}^{\dagger}_{j,V}\hat{a}_{j,V}$ and $\hat{N}_{bj,-}=\hat{b}^{\dagger}_{j,H}\hat{b}_{j,H}-\hat{b}^{\dagger}_{j,V}\hat{b}_{j,V}$ $(j=2,3)$, respectively. The subscripts $j=2,3$ represent the operators corresponding to the 'Chiral matter-photon interaction' and 'Polarization detection' modules, respectively. Under the first-order approximation, the annihilation operator $\hat{a}$ can be expressed as $\hat{a}=\alpha+\delta\hat{a}$, where $\alpha$ represents a real-valued classical amplitude and $\delta\hat{a}$ corresponds to the quantum fluctuation. Therefore, the operators $\hat{N}_{a2,-}$ and $\hat{N}_{b2,-}$ can be written as:

$$\begin{aligned}
\hat{N}_{a2,-} &= \hat{a}^{\dagger}_{2,H}\hat{a}_{2,H}-\hat{a}^{\dagger}_{2,V}\hat{a}_{2,V} \\
&= \left(\alpha_H+\delta\hat{a}^{\dagger}_{2,H}\right)\left(\alpha_H+\delta\hat{a}_{2,H}\right)-\left(\alpha_V+\delta\hat{a}^{\dagger}_{2,V}\right)\left(\alpha_V+\delta\hat{a}_{2,V}\right) \\
&= \alpha\left(\delta\hat{a}^{\dagger}_{2,H}-\delta\hat{a}^{\dagger}_{2,V}+\delta\hat{a}_{2,H}-\delta\hat{a}_{2,V}\right) \\
&= \alpha\left(\delta\hat{a}^{\dagger}_{2,H-V}+\delta\hat{a}_{2,H-V}\right)=\alpha\delta\hat{X}_{a,H-V},
\end{aligned} \tag{3}$$

and $\hat{N}_{b2,-}=\alpha\delta\hat{X}_{b,H-V}$, respectively. In deriving above expressions, we define the quantum fluctuation $\delta\hat{a}_{2,H-V}\equiv\delta\hat{a}_{2,H}-\delta\hat{a}_{2,V}$, and set an equal classical amplitude in the two polarizations, i.e. $\alpha_H=\alpha_V=\alpha$. This CV polarization entangled state manifests as quantum correlations between Stokes operators of modes $a$ and $b$ (*70*, *71*), where the quantum noise of difference between $\hat{N}_{a2,-}$ and $\hat{N}_{b2,-}$, i.e. $\hat{N}_{2.-}=\hat{N}_{b2.-}-\hat{N}_{a2.-}=\alpha(\delta\hat{X}_{b,H-V}-\delta\hat{X}_{a,H-V})$, is suppressed by a factor of $\sqrt{G^2+g^2}$ below the shot noise level (see Methods), thereby rendering the entangled probe a sensitive probe for chiral discrimination.

The chiral matter-photon interaction is a circular birefringence process in physics, which causes left- and right-circularly polarized light (LCP and RCP) to acquire different phases. In this sense, the linearly polarized modes $\{\hat{a}_{2,H},\hat{a}_{2,V},\hat{b}_{2,H},\hat{b}_{2,V}\}$ are decomposed into circularly polarized modes $\{\hat{a}_{2,R},\hat{a}_{2,L},\hat{b}_{2,R},\hat{b}_{2,L}\}$, following

$$\begin{pmatrix}\hat{a}_{2,R}\\ \hat{a}_{2,L}\end{pmatrix}=M\begin{pmatrix}\hat{a}_{2,H}\\ \hat{a}_{2,V}\end{pmatrix},\ \begin{pmatrix}\hat{b}_{2,R}\\ \hat{b}_{2,L}\end{pmatrix}=M\begin{pmatrix}\hat{b}_{2,H}\\ \hat{b}_{2,V}\end{pmatrix},\ M=\frac{1}{\sqrt{2}}\begin{pmatrix}1 & -i\\ 1 & i\end{pmatrix}. \tag{4}$$

These circularly polarized modes experience phase change $\theta$ through the chiral sample and evolve into $e^{i\theta}\{\hat{a}_{2,R},\hat{b}_{2,R}\}$ and $e^{-i\theta}\{\hat{a}_{2,L},\hat{b}_{2,L}\}$. Eventually, as illustrated in Fig. 1, this leads to a small polarization rotation in the entangled probe, characterized by the linearly polarized modes $\{\hat{a}_{3,H}, \hat{a}_{3,V}, \hat{b}_{3,H}, \hat{b}_{3,V}\}$ with

$$\begin{pmatrix} \hat{a}_{3,H} \\ \hat{a}_{3,V} \end{pmatrix} = M^{-1} \begin{pmatrix} e^{i\theta}\hat{a}_{3,R} \\ e^{-i\theta}\hat{a}_{3,L} \end{pmatrix}, \begin{pmatrix} \hat{b}_{3,H} \\ \hat{b}_{3,V} \end{pmatrix} = M^{-1} \begin{pmatrix} e^{i\theta}\hat{b}_{3,R} \\ e^{-i\theta}\hat{b}_{3,L} \end{pmatrix}. \tag{5}$$

The four output modes $\{\hat{a}_{3,H}, \hat{a}_{3,V}, \hat{b}_{3,H}, \hat{b}_{3,V}\}$ are directed into two balanced amplified photodetectors (Thorlabs, PDB450A). The polarization differential signal $\langle \hat{N}_{3,-} \rangle$ is measured in the 'Polarization detection' module as shown in Fig. 1, where the operator $\hat{N}_{3,-}$ takes the expression

$$\hat{N}_{3,-} = \hat{N}_{b3,-} - \hat{N}_{a3,-} = 4\alpha^2 \sin(2\theta) + \hat{N}_{2,-} \approx 8\alpha^2\theta + \alpha\left(\delta\hat{X}_{b,H-V} - \delta\hat{X}_{a,H-V}\right) \tag{6}$$

which serves as the measurement observable for chiral discrimination. The first term in Eq. (6) corresponds to the chiral signal, scaling linearly with the polarization rotation angle $\theta$ in the small-angle regime, while the second term denotes the quantum noise, which is suppressed due to polarization entanglement between modes $a$ and $b$. According to the error propagation formula, the ultimate measurement sensitivity is given by:

$$\delta\theta = \frac{\Delta\hat{N}_{3,-}}{\partial\langle\hat{N}_{3,-}\rangle / \partial\theta} = \frac{1}{2\sqrt{G^2+g^2}\sqrt{N_t}} \tag{7}$$

where the variance $\Delta\hat{N}_{3,-} = \sqrt{\langle\hat{N}_{3,-}^2\rangle - \langle\hat{N}_{3,-}\rangle^2}$ characterize the noise of the observable $\hat{N}_{3,-}$, $N_t$ represents the total photon number probing the chiral sample. This quantum enhancement originates from the suppression of quantum fluctuations of differential signal $\hat{N}_{3,-}$, whose variance is same as $\hat{N}_{2,-}$ given by $N_t \big/ \sqrt{G^2+g^2}$ (See Methods).

**Characterization of Quantum Enhancement:**

A bright, high-quality CV entangled source is the prerequisite for chiral discrimination. To this end, we seed a weak coherent state into each PA to enhance the generation rate of polarization-entangled states. The correlated photon generation rate reaches $10^{15}$ Hz, which is approximately $10^{10}$ times higher than that of typical two-photon N00N states (*45*, *75*, *76*). In the absence of chiral molecules, we measure the observable $\hat{N}_{3,-} = \hat{N}_{b3,-} - \hat{N}_{a3,-}$ to compare the noise spectra of the coherent and entangled probes under identical photon flux. The measured noise spectrum of the coherent probe serves as the shot-noise level, representing the SNL in the absence of quantum entanglement. As shown in Fig. 2A, the CV-entangled probe suppresses quantum noise by approximately 6 dB relative to the shot noise level.

We next characterize quantum enhancement of signal-to-noise ratio (SNR) for polarization measurement with signal modulation. To circumvent dominant low-frequency technical noise, we employ a signal modulation scheme that shifts the polarization-rotation signal to a higher frequency which is commonly employed in sensitive chiral measurement (*7*). Experimentally, we apply a 1.5 MHz sinusoidal modulation to the polarization of modes $a$ and $b$. After interaction with the test sample, the resulting polarization rotation signal is expressed as

$$\left\langle \hat{N}_{3,-} \right\rangle = N_t \sin 2\theta \sin(A \sin \omega t) \propto J_1(A) N_t \theta \sin \omega t, \quad (8)$$

where $J_1(A)$ denotes the first-order Bessel function and $A$ is the modulation depth. The measured signal is encoded in the amplitude of the modulated carrier wave. As shown in Fig. 2B, the peak is located at the carrier frequency, and its signal amplitude directly reflects the polarization rotation angle $\theta$. The differnence betweeen the signal amplitude and background noise level is defined as SNR. The entangled probe achieves an approximately 5 dB improvement in SNR compared with the coherent probe.

To precisely determine the chiral molecular polarization rotation angle $\theta$, we calibrate the relationship between $\theta$ and detected signal amplitude (see Methods) with a half-wave plate (HWP), which provides controllable polarization rotation angles. As illustrated in Fig. 2C, the SNR enhancement remains nearly constant across different polarization rotation angles, highlighting the robustness of our setup for chiral discrimination.

**Quantum-elevated discrimination of amino acids:**

As fundamental building blocks of life, amino acids play central roles in metabolism, neurotransmission, and hormone production. The ability to precisely discriminate their chirality is essential for applications in nutrition, medical diagnostics, pharmaceuticals, and food quality control. In this work, we employ arginine as an example to demonstrate quantum-elevated chiral discrimination and enantiomeric excess (*e.e.*) measurement. Arginine is particularly important due to its involvement in nitric oxide production, cardiovascular function, and the urea cycle, making its accurate quantification critical for assessing metabolic health and diagnosing related disorders.

Under liquid-phase conditions that mimic typical biological environments, we measure the concentrations of pure L- and D-arginine using coherent and entangled probes, each operating under identical photon flux. The aqueous amino acid solutions are contained in a 35 mm anti-reflection-coated glass cell mounted in the optical path of the chiral probe. Owing to reduced quantum fluctuations, the entangled probe exhibits enhanced sensitivity in resolving both the magnitude and sign of polarization rotation compared with the classical probe of coherent light (left panels of Fig. 3). As demonstrated in the upper panels of Fig. 3A and Fig. 3B, the shaded regions around the mean values represent measurement uncertainty. The narrower spread for the entangled probe highlights its superior resolving capability compared to the coherent probe. Amino acid-chiral probe interactions induce polarization rotation, which is measured via correlation detection and Fourier analysis. The signal amplitude at the 1.5 MHz modulation frequency scales approximately linearly with arginine concentration. As illustrated in the lower panels of Fig. 3A and Fig. 3B, the optical rotation signal gradually diminishes with decreasing chiral solution concentration and eventually falls below the shot-noise level. Below a concentration threshold of approximately 0.075 g/ml, the signal can only be resolved using CV entangled probe. Experimentally, the minimum resolvable concentration of arginine with the CV entangled probe is about three times lower than that achievable with a coherent probe. While the signal amplitude obtained from the spectrum analyzer reflects concentration, it does not distinguish the handedness of chiral enantiomers. To determine enantiomeric handedness, we exploit the formal analogy between the unitary transformation induced by a chiral sample and that of a HWP. By rotating the HWP's optical axis clockwise or counterclockwise to nullify the signal, the direction of rotation indicates the left- or right-handedness of the enantiomers. Alternatively, a lock-in amplifier can be employed to directly extract both the polarity and amplitude of the signal. Both approaches enable unambiguous discrimination of chiral enantiomers.

In practical research, chiral samples are often mixtures of different enantiomers. Determining the *e.e.* is crucial for evaluating the efficacy and safety of pharmaceuticals (*77*), assessing asymmetric synthesis (*78*), monitoring spontaneous deracemizations (*79*), and investigating the origin of homochirality (*80*, *81*). Accurate measurement of *e.e.* is therefore essential for characterizing the properties and functions of chiral compounds in both scientific and industrial contexts. Enantiomeric excess, defined as $([L]-[D])/([L]+[D])\times 100\%$, quantifies the relative abundance of L- and D-enantiomers, where $[L]$ and $[D]$ represent the concentrations of the left- and right-handed forms, respectively. We evaluate *e.e.* by preparing aqueous mixtures of L- and D-arginine using the same polarization-based chiral discrimination method. At high *e.e.* of 90% in Fig. 4, the dominant enantiomer reaches concentrations up to 0.12 g/mL. As shown in Fig. 4, CV entangled probes resolve lower *e.e.* values beyond the SNL, approaching the quantum-enhanced noise level. The minimum resolvable *e.e.* achieved with the entangled probe (red star in Fig. 4B) is significantly smaller than that of the coherent probe (black star in Fig. 4A).

## DISCUSSION

In this study, we demonstrate a quantum metrological approach to chiral discrimination, leveraging CV polarization-entangled states generated by a pair of PAs implemented $^{85}$Rb atomic ensembles. Unlike discrete-variable entanglement employed in previous quantum sensing experiments (*40–45*), our method offers enhanced squeezing performance and scalability to high photon numbers, mitigating the tradeoff between enhancing sensitivity and minimizing optical damage. With an identical sensing photon number, our quantum protocol surpasses the SNL by 5 dB, enabling enantiomer discrimination at concentrations up to three times lower than those detectable with conventional laser-based chiroptical techniques. The slight reduction from the 6 dB suppression of the entangled source is attributed to unavoidable transmission losses.

Our quantum-enhanced approach marks a fundamental advancement in chiral sensing. Conventional methods, such as micro-nano platforms engineering (*19–24*), multi-pass strategy with cavity (*25–27*), and temporal or spatial light tailoring (*31–39*), improve sensitivity primarily by strengthening chiral matter-light interactions, yet increase the risk of optical damage. In contrast, our method enhances sensitivity by suppressing the inherent quantum noise of the optical probe below the shot-noise limit with a moderate photon flux, offering a safer alternative for probing light-sensitive samples. The future integration of a quantum-entangled light source into conventional methods presents a promising path to achieving unprecedented sensitivity levels. Such integration paves the way for high-precision chiral analysis in biology, chemistry, and pharmaceutical sciences.

## MATERIALS AND METHODS

### Experimental Implementation

As shown in Fig. 5, the experimental implementation is constituted by three procedures---'Entangled State Preparation', 'Phase Stabilization', and 'Chiral Quantum Discrimination'. In the 'Entangled State Preparation' module, we prepare two mode squeezed state (TMSS) via a stimulated four-wave-mixing (FWM) process of $^{85}$Rb atom, where a frequency-degenerate pump generates correlated signal and idler beams. The pump beam is generated by a Ti: Sapphire laser (Sirah Matisse 2 TS), tuned approximately 1 GHz to the blue of the $^{85}$Rb $5S_{1/2}$, $F=2$ to $5P_{1/2}$, $F'=2$ transition. A small portion of this blue-tuned laser is split using a polarized beam splitter (PBS) to serve as the seed of the signal beam, enhancing the FWM process. This seed beam is then red-shifted by approximately 3.04 GHz relative to the pump beam using an acousto-optic modulator (AOM) in a double-pass configuration. The Faraday rotator is used to rotate the seed polarization from horizontal to vertical. Both the pump and seed beams are subsequently split by a half-wave plate (HWP) and a PBS into two channels, referred to as the upper and lower channels, which are separated by a distance of approximately 4 mm. In each channel, the pump beam has a power of 200 mW with vertical polarization, and the seed beam has a power of 25 μW with horizontal polarization. Subsequently, they are combined through a Glan-Laser polarizer at a crossing angle of 0.3° and focused at the center of the $^{85}$Rb atomic vapor cell to generate a pair of TMSS in two channels.

The $^{85}$Rb atomic vapor cell (12.5 mm long) is heated to 105°C with the high transmission efficiency 90%. After passing through the atomic vapor cell, the correlated signals (amplified seeds) and idlers are generated in pairs, with the signals red-shifted and the idlers blue-shifted by 3.04 GHz relative to the pump beam. The frequency difference between the signal and idler beams is approximately 6.08 GHz. The frequency difference between two modes is so small that their interaction with the chiral sample can be regarded as identical. Therefore, the frequency non-degeneracy does not play any role in the chiral measurement. To filter out the excess pump beams, a Glan-Thomson polarizer (GT) with an extinction ratio of $10^5$:1 is utilized. The polarization of the signal and idler beams for the lower channel is adjusted to vertical polarization via HWPs. A pair of TMSSs with orthogonal polarization is then combined using a PBS to generate a polarization-entangled state. The optical power of the entangled probe is 500 $\mu$W , corresponding to a photon flux of $2\times10^{15}$ photons/second. Over 99% of the optical intensity of the TMSS output emerges from the right side of the PBS, with a mode matching efficiency exceeding 95%, as measured by observing interference fringes. Meanwhile, the residual 1% of the output beam, from the other side of the PBS, is directed into the 'Phase Stabilization' module. This residual light is utilized to stabilize the relative phase between the horizontal and vertical components of modes $a$ and $b$ . A detailed theoretical analysis is provided in the Supplementary Materials.

In the 'Chiral Quantum Discrimination' module, the entangled probe traverses a chiral solution cell coated with anti-reflection films on both sides to minimize optical loss and back reflection. The correlated signal and idler beams are subsequently separated by a PBS into horizontal and vertical polarization components and directed into two balanced photodetectors (PDBs, Thorlabs PDB450A) to measure the polarization intensity difference. The two PDBs are equipped with high quantum efficiency (96%) photodiodes (S3883) to minimize detection losses. A differential electronic circuit is employed to directly measure the intensity difference between the two PDB signals. The differential signal is sent to a spectrum analyzer to retrieve the polarization rotation signal. The two digital storage oscilloscopes are used to monitoring the stability of the whole setup. In the absence of the chiral sample, the noise spectra of the polarization-entangled states are measured over a 6 MHz frequency range, as shown in Fig. 2a of the main text, with a maximum squeezing of 6 dB. By placing the chiral amino acids in the optical path, both chiral discrimination and enantiomeric excess can be measured, as shown in Figs. 3 and 4, respectively.

**Quantum Fluctuation Suppression in CV Polarization Entanglement**

According to Eq. (1), the input-output relations of the PA can be expressed as follows:

$$\begin{aligned}
\hat{a}_{1,H} &= G\hat{a}_{0,H} + g\hat{b}^{\dagger}_{0,H}, \\
\hat{b}_{1,H} &= G\hat{b}_{0,H} + g\hat{a}^{\dagger}_{0,H}, \\
\hat{a}_{1,V} &= G\hat{a}_{0,V} + g\hat{b}^{\dagger}_{0,V}, \\
\hat{b}_{1,V} &= G\hat{b}_{0,V} + g\hat{a}^{\dagger}_{0,V}.
\end{aligned} \tag{9}$$

The horizontal and vertical polarization fields of modes $a$ and $b$ are combined using respective PBSs to generate a CV polarization entangled state. We define $\hat{a}_{2,H} = \hat{a}_{1,H}$, $\hat{a}_{2,V} = -\hat{a}_{1,V}$, $\hat{b}_{2,H} = \hat{b}_{1,H}$ and $\hat{b}_{2,V} = \hat{b}_{1,V}$. The entanglement manifests the correlation between the Stokes operators $\hat{N}_{a2,-} = \hat{a}^{\dagger}_{2,H}\hat{a}_{2,H} - \hat{a}^{\dagger}_{2,V}\hat{a}_{2,V}$ and $\hat{N}_{b2,-} = \hat{b}^{\dagger}_{2,H}\hat{b}_{2,H} - \hat{b}^{\dagger}_{2,V}\hat{b}_{2,V}$ of modes $a$ and $b$, leading to the suppression of quantum fluctuation in their difference, $\hat{N}_{2,-} = \hat{N}_{b2,-} - \hat{N}_{a2,-}$. The variance of $\hat{N}_{a,-}$ can be derived as

$$\Delta^2 \hat{N}_{2,-} = \left\langle \hat{N}^2_{2,-} \right\rangle - \left\langle \hat{N}_{2,-} \right\rangle^2 = \frac{N_t}{G^2 + g^2}, \tag{10}$$

where $\hat{N}_t$ denotes the total photon numbers of the entangled state. The fluctuation is suppressed by a factor of $G^2 + g^2$ below the shot noise level. We now present a physical explanation of the quantum noise suppression. In the stimulated four-wave mixing, a seed with intensity $|\beta|^2$ is amplified into a signal $\left(G^2|\beta|^2\right)$ and an idler $\left(g^2|\beta|^2\right)$. For the intensity difference observable, the noise for uncorrelated coherent fields would be $\left(G^2 + g^2\right)|\beta|^2$. In contrast, the quantum correlations in the two-mode squeezed state suppress this noise to the seed's level, $|\beta|^2$, corresponding to a noise suppression factor of $G^2 + g^2$. Consequently, the sensitivity, scaling with the noise standard deviation, is enhanced by a factor of $\sqrt{G^2 + g^2}$.

This entangled state serves as the probe to perform chiral measurement. Due to the circular birefringence effect, the output polarization of mode $a$ undergoes a slight rotation characterized by

$$\begin{pmatrix} \hat{a}_{3,H} \\ \hat{a}_{3,V} \end{pmatrix} = \begin{pmatrix} \cos\theta & \sin\theta \\ -\sin\theta & \cos\theta \end{pmatrix} \begin{pmatrix} \hat{a}_{2,H} \\ \hat{a}_{2,V} \end{pmatrix}, \tag{11}$$

which is identical for mode $b$. Consequently, the expression of $\hat{N}_{3,-}$ is derived as shown in Eq. (6). The quantum noise fluctuation remains the same as $\hat{N}_{2,-}$, yielding sensitivity enhancement of $\sqrt{G^2 + g^2}$ below SNL. The quantum noise suppression is illustrated in Fig. 1 with the phase diagram of $\delta\hat{X}_-$ and $\delta\hat{X}_+$, which are squeezed and anti-squeezed, respectively. The fluctuation of $\delta\hat{X}_-$ represent the noise properties of both $\hat{N}_{2,-}$ and $\hat{N}_{3,-}$.

**Data Calibration and Processing**

We employ a HWP to calibrate the relationship between polarization-rotation angle and signal amplitude. The rotation angle is referenced using a high-precision rotation stage, while the signal amplitude at the carrier frequency is acquired with a spectrum analyzer. A calibration function is obtained by fitting the measured data, enabling quantitative conversion between signal amplitude and polarization-rotation angle. This calibration is then used to infer the polarization-rotation angles arising from chiral molecules based on the measured signal amplitude. All rotation angles reported for amino acids in Figs. 3-4 are obtained using this method.

## Acknowledgments:

**Funding:**

Quantum Science and Technology-National Science and Technology Major Project grant 2021ZD0303200

Quantum Science and Technology-National Science and Technology Major Project grant 2024ZD0302200

National Natural Science Foundation of China grant 12234014

National Natural Science Foundation of China grant 12204303

National Natural Science Foundation of China grant 12374331

National Natural Science Foundation of China grant 11904227

National Natural Science Foundation of China grant 11654005

National Natural Science Foundation of China grant 12404416

National Natural Science Foundation of China grant 12504412

Fundamental Research Funds for the Central Universities

Shanghai Municipal Science and Technology Major Project grant 2019SHZDZX01

Shanghai Science and Technology Innovation Action Plan grant 24LZ1401400

Shanghai Science and Technology Innovation Action Plan grant 24ZR1437900

Innovation Program of Shanghai Municipal Education Commission grant 202101070008E00099

National Key Research and Development Program of China grant 2016YFA0302001

Fellowship of China Postdoctoral Science Foundation grant GZB20230424

State Key Laboratory of IPOC (BUPT) grant IPOC2025ZT04

Shanghai Talent Program (WPZ)

**Author contributions:**

Conceptualization: Y.Q.Y., G.Z.B., W.P.Z.

Methodology: Y.Q.Y., X.L.H, W.D., S.H.W., P.Y.Y., G.Z.B.

Investigation: Y.Q.Y., X.L.H, W.D., S.H.W., P.Y.Y., G.Z.B.

Visualization: Y.Q.Y., X.L.H.

Funding acquisition: Y.Q.Y., W.D., P.Y.Y., S.H.W., G.Z.B., W.P.Z.

Project administration: Y.Q.Y., G.Z.B., W.P.Z.

Supervision: G.Z.B., W.P.Z.

Writing – original draft: Y.Y.Q., X.L.H, G.Z.B., W.P.Z.

Writing – review & editing:

**Competing interests:** Authors declare that they have no competing interests.

**Data and materials availability:** All data are available in the main text or the supplementary materials.

## Supplementary Materials

Performance of two-mode squeezed state under loss

Performance of continuous-variable entangled state in chiral sensing under lossy conditions

Phase stabilization

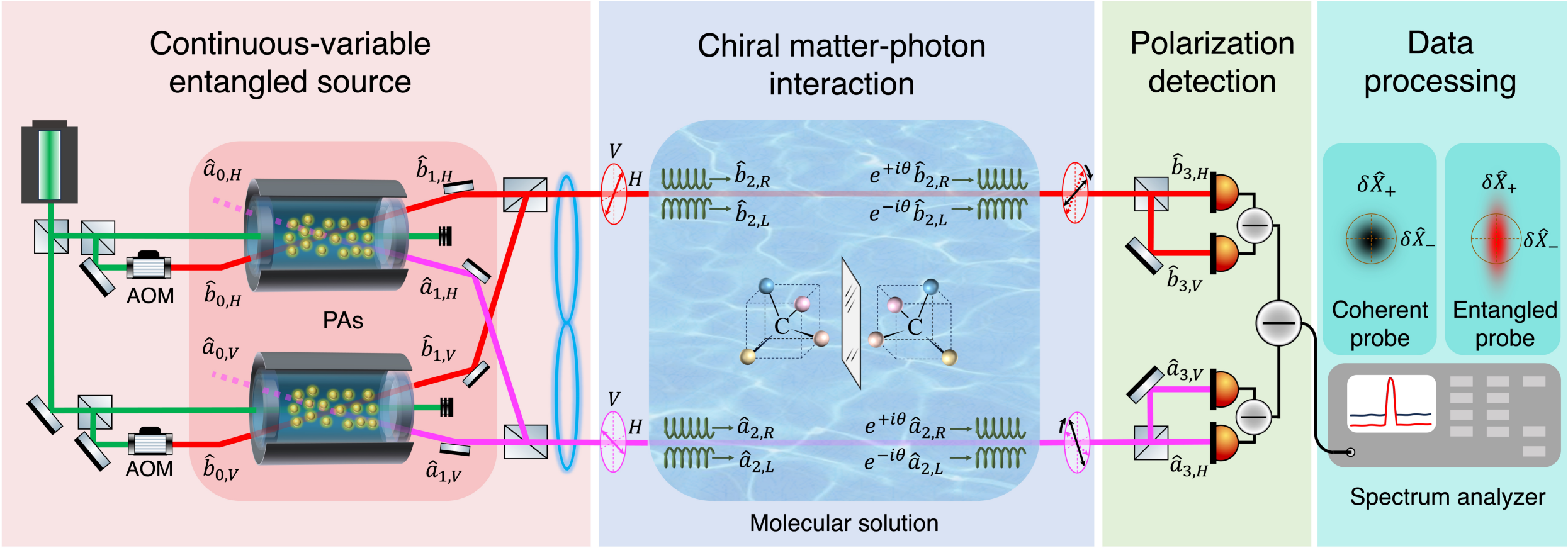

**Fig. 1. Schematic Diagram of Quantum-elevated Chiral Discrimination.** In the 'Continuous-variable entangled source', the continuous-variable entangled state is created by superposing a pair of two-mode squeezed state generated by stimulated four-wave mixing process in two parametric amplifiers (PAs). The PAs are implemented using $^{85}$Rb atomic ensemble, seeded with red-tuned light from an acousto-optic modulator (AOM). In the 'Chiral matter-photon interaction', due to the circular birefringence effect, the left and right circularly polarized lights (LCP and RCP) experience different phases, resulting into either clockwise or counterclockwise polarization rotation upon interaction with L-amino acids or D-amino acids, respectively. The positivity and magnitude of this rotation angle serve as reliable indicators of handedness and concentration of chiral molecules. In the 'Polarization detection', the polarization differential signal $\langle \hat{N}_{3,-} \rangle = \langle \hat{N}_{b3,-} - \hat{N}_{a3,-} \rangle$ is measured. In the 'Data Processing', a spectrum analyzer is employed to analyze the polarization differential signal and retrieve the polarization rotation angle $\theta$. As demonstrated in the above panels, the quantum fluctuation of $\delta \hat{X}_-$ is suppressed compared to the coherent probe, resulting in quantum-enhanced chiral discrimination. The operators $\delta \hat{X}_-$ and $\delta \hat{X}_+$ are defined as $\delta \hat{X}_- = \delta \hat{X}_{b,H-V} - \delta \hat{X}_{a,H-V} = (\delta \hat{b}^{\dagger}_{3,H-V} + \delta \hat{b}_{3,H-V}) - (\delta \hat{a}^{\dagger}_{3,H-V} + \delta \hat{a}_{3,H-V})$ and $\delta \hat{X}_+ = \delta \hat{X}_{b,H+V} + \delta \hat{X}_{a,H+V} = (\delta \hat{b}^{\dagger}_{3,H+V} + \delta \hat{b}_{3,H+V}) + (\delta \hat{a}^{\dagger}_{3,H+V} + \delta \hat{a}_{3,H+V})$, respectively.

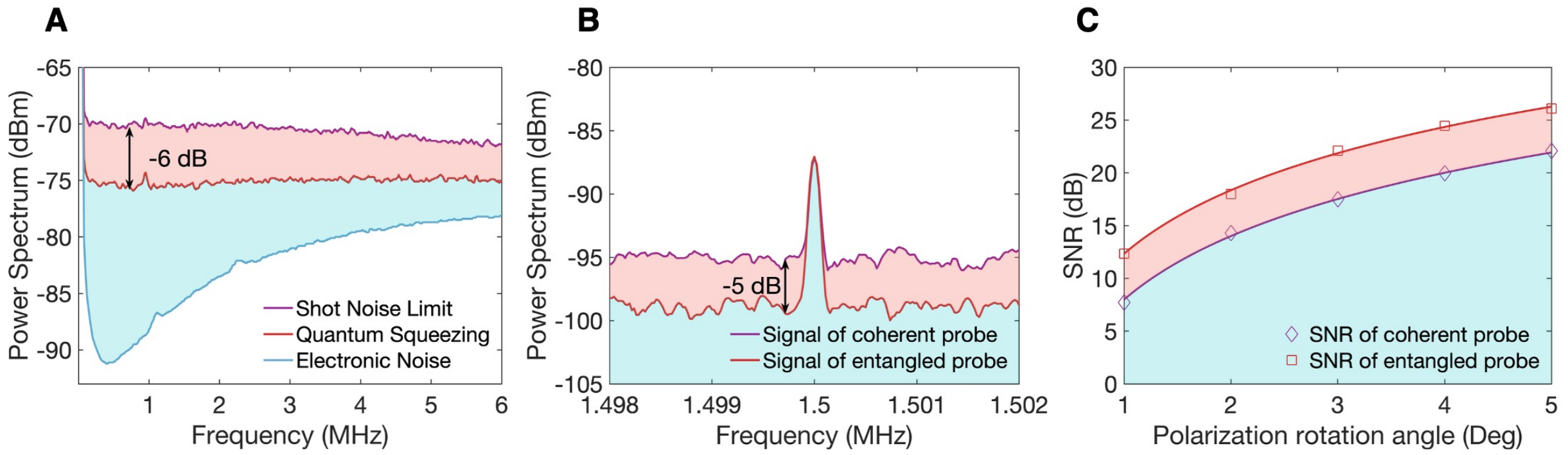

**Fig. 2. Noise and signal-to-noise ratio (SNR) comparison between coherent and entangled probe.** (**A**) Noise power spectrum measured by spectral analyzer. The resolution bandwidth (RBW) and video bandwidth (VBW) are 30 kHz and 300 Hz, respectively. The balanced photodetector's (PDB) transimpedance gain is $10^5$ V/A with a bandwidth of 4 MHz. The purple curve denotes the shot noise limit of coherent state. The red curve represents the intensity-difference squeezing of the entangled state, reaching $\sim 6$ dB of squeezing at 0.7 MHz. The spike near 1 MHz is due to the inherent noise of pump laser. The cyan curve denotes the electronic noise level of PDB. (**B**) Measured signal and noise spectra of the classical probe (purple) and quantum probe (red) for a half-wave plate (HWP) with its optical axis oriented at $0.5^{\circ}$ relative to the horizontal direction. The entangled probe achieves a $\sim 5$ dB enhancement in SNR compared to classical probes. (**C**) SNR comparison between the classical (purple) and quantum (red) probe for different polarization rotation angles. The RBW and VBW are set to 100 Hz and 1 Hz, respectively, for data collection in (B) and (C).

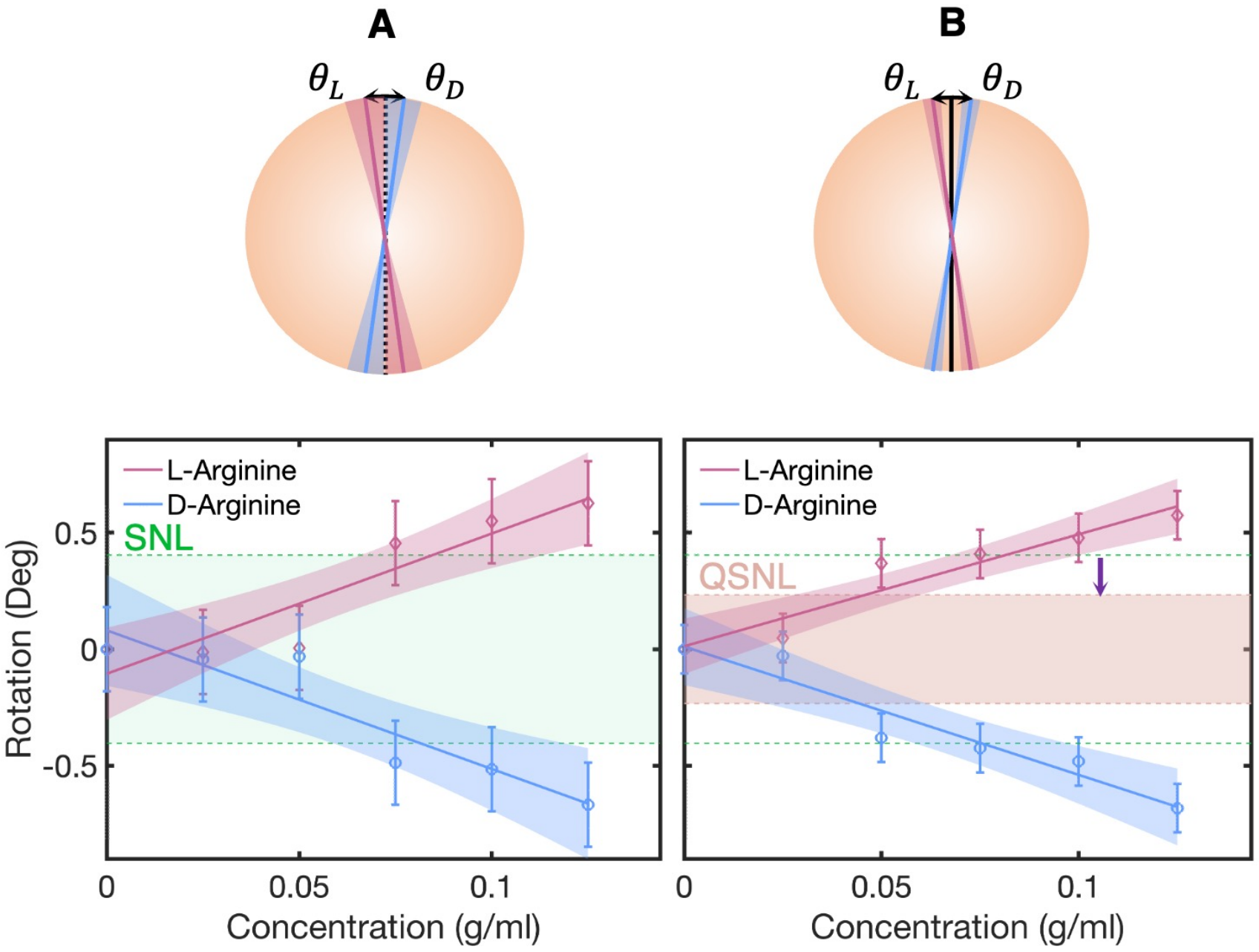

**Fig. 3. Chiral discrimination of L/D-amino acids with coherent and entangled probe.** Under liquid-phase conditions, we measure L- and D-arginine at varying concentrations using both coherent and quantum-entangled probes. The shot-noise level (SNL) defines the minimum resolvable polarization rotation angle for coherent light, while the quantum-squeezed noise level (QSNL) represents the rotation angle achievable with entangled light due to suppressed quantum fluctuations. (**A**) The coherent probe exhibits reduced sensitivity in detecting small rotation angles compared to the entangled case. (**B**) At equivalent photon flux, the entangled probe enables improved resolution of polarization rotations $\theta_L$ and $\theta_D$, corresponding to left- and right-handed enantiomers, respectively. All data represent the average of five independent measurements. Shaded regions of the fitted linear curves represent 95% confidence intervals.

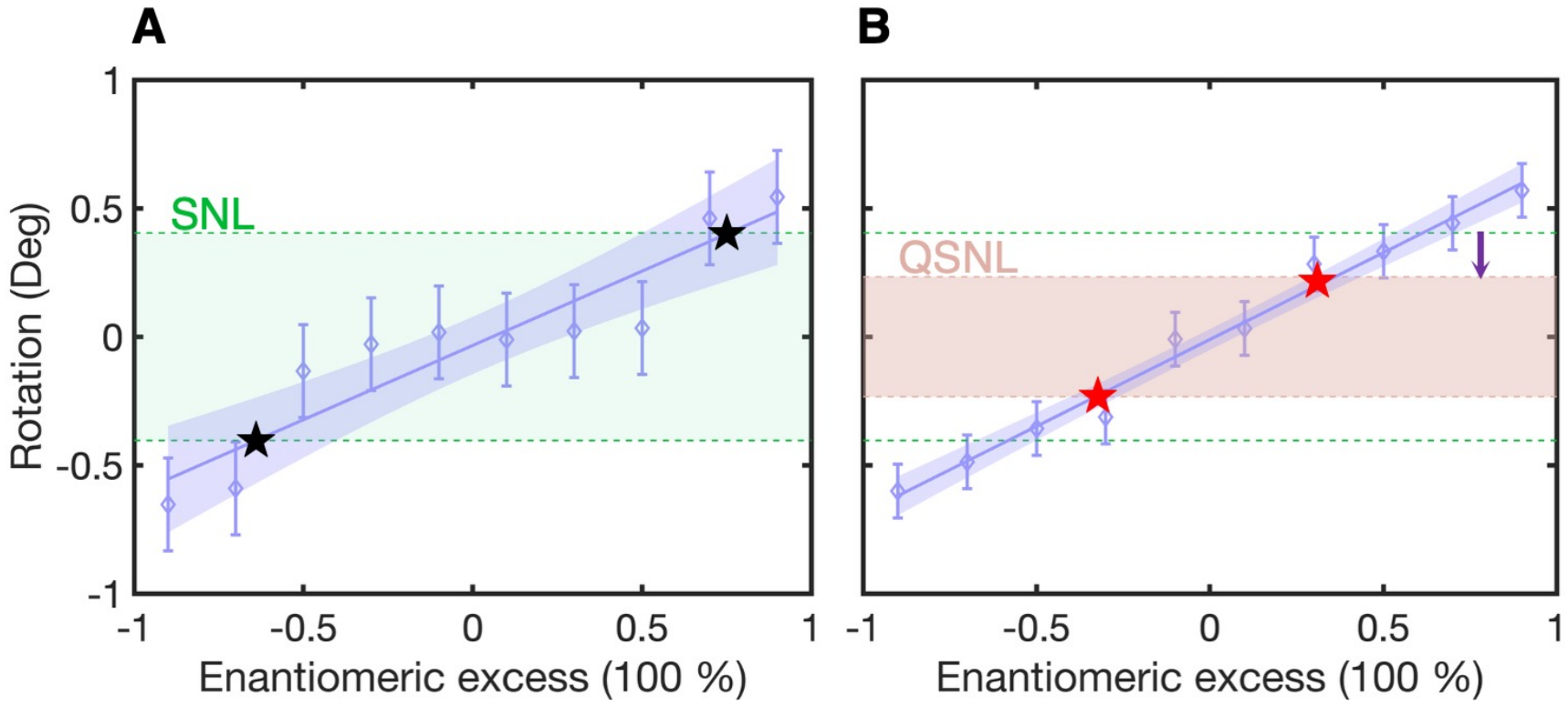

**Fig. 4. Enantiomeric excess determination of mixed L/D-amino acids with coherent and entangled probe.** Under liquid-phase conditions, we measure the enantiomeric excess (*e.e.*) of L/D-arginine mixtures at varying concentrations using both coherent (**A**) and quantum-entangled (**B**) probes. The shot-noise level (SNL) and quantum-squeezed noise level (QSNL) determine the minimum resolvable *e.e.* achievable with coherent and entangled probes, respectively, as indicated by black and red star. At equal photon flux, the entangled probe exhibits enhanced sensitivity, enabling more precise discrimination of *e.e.*, corresponding to the imbalance between left- and right-handed enantiomers. All data represent the average of five independent measurements. Shaded regions of the fitted linear curves represent 95% confidence intervals.

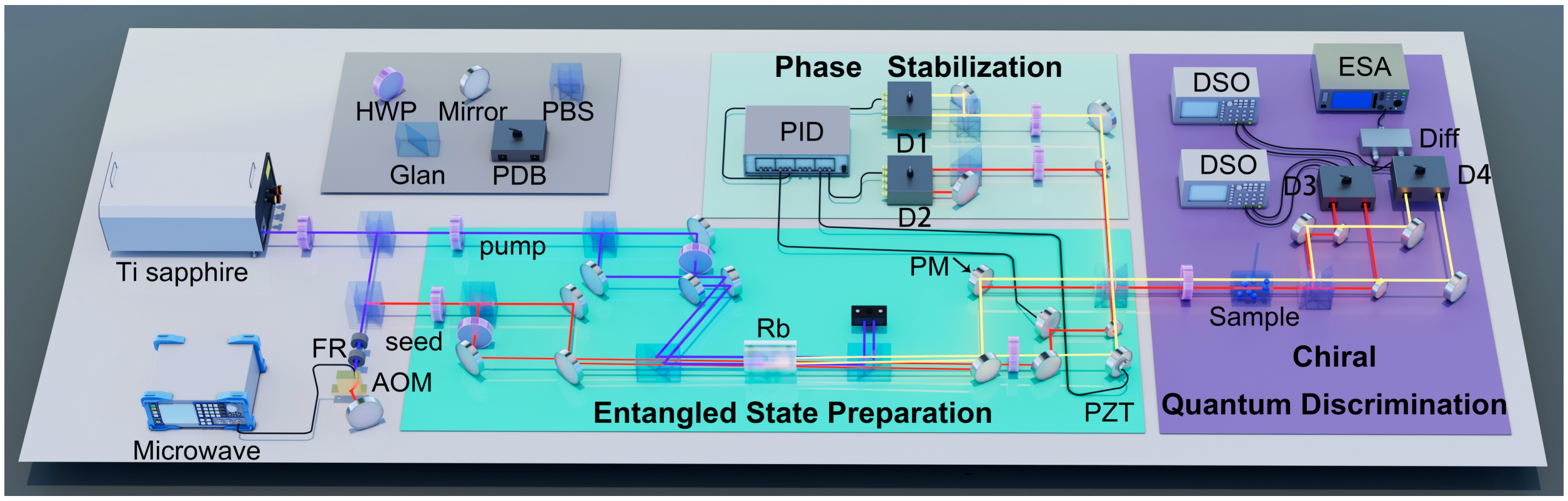

**Fig. 5 Experimental setup.** Our experimental setup comprises three main modules: ‘Entangled state preparation’, ‘Phase Stabilization’, and ‘Chiral Quantum Discrimination’. In the ‘Entangled State Preparation’ module, a continuous-variable (CV) polarization-entangled state is generated. The microwave source generates a 1.518 GHz signal that drives the AOM to tune the frequency of the seed light. An atomic ensemble of rubidium (Rb) atoms is employed to generate correlated signal-idler beams through a four-wave mixing (FWM) process based on Raman interactions. Two coherent FWM processes are superimposed to produce the entangled state. Driving the PZT (indicated by the black arrow) with a sinusoidal signal modulates the H-V phase difference, thereby realizing polarization modulation (PM). The ‘Phase Stabilization’ module is designed to stabilize the phase of the CV entangled state. Error signals are generated by detectors D1 and D2 and are fed into a PID controller. The PID controller outputs modulation signals to drive PZT and maintain phase stability. The ‘Chiral Quantum Discrimination’ is utilized for chiral discrimination. Intensity difference measurements between the horizontal and vertical polarization components of the signal and idler beams are conducted by detectors D3 and D4, respectively. The electronic signals from these detectors are sent to a Diff circuit, which subtracts the two signals and outputs the result to an electrical spectrum analyzer (ESA) for spectral analysis. Digital storage oscilloscopes (DSOs) monitor the stability of the entire setup. The detectors D1-D4 are balanced amplified photodetectors (PDB). GT: Glan lens. D1-D4: Balanced Amplified Photodetectors. AOM: acoustic optical modulator. FR: Faraday Rotator. PZT: piezoelectric transducer. PM: Polarization modulation. ESA: Electronic Spectral analyzer. DSO: Digital Signal Oscilloscope. PID: Proportional-Integral-Derivative control systems. Diff: difference channel.